\documentclass{moriond}
\usepackage[numbers]{natbib}

\newcommand{\Photo}{}

\begin{document}
\vspace*{4cm}
\title{Highlights from TeV Extragalactic Sources} 

\author{ Elisa Prandini }

\address{Department of Physics and Astronomy, Via Marzolo 8,\\
35131 Padova, Italy}

\maketitle\abstracts{
The number of discovered TeV sources populating the extragalactic sky in 2017 is nearly 70, mostly blazars located up to a redshift $\sim$1. Ten years ago, in 2007, less than 20 TeV emitters were known, up to a maximum redshift of 0.2. 
This is a major achievement of current generation of Cherenkov telescopes operating in synergy with optical, X-ray, and GeV gamma-ray telescopes. A review of selected results from the extragalactic TeV sky is presented, with particular emphasis on recently detected distant sources.
}

\section{Introduction}\label{sec:intro}
Very high energy gamma rays (VHE; E $>$ 100\,GeV) are messengers of the most energetic phenomena in the Universe. A number of pulsars, pulsar wind nebulae, supernova remnants and micro-quasars in our own Galaxy have been detected to emit such energetic radiation. Outside the Milky Way, VHE gamma rays have been seen from ultra-relativistic jets of particles escaping super-massive black holes and from galaxies with an exceptional rate of star formation. The VHE sky counts nowadays more than 150 sources.  A large fraction of these sources have been detected for the fist time at VHE by the current generation of Imaging Atmospheric Cherenkov Telescopes (IACTs), namely H.E.S.S., MAGIC, and VERITAS. IACTs are designed to detect the faint Cherenkov flashes that originate from a particle shower that is formed when an energetic photon impacts the Earth atmosphere. 
 
All  three telescope systems use large reflective mirrors to focus the Cherenkov (near-UV and optical) light into meter-size cameras pixelized by photomultiplier tubes. The typical signature of a gamma-ray induced shower in the camera is an image of elliptical shape and few $ns$ duration, that has to be separated from the dominating background of mainly proton-induced showers. 

The H.E.S.S. array of telescopes \cite{Hinton04} is located in Khomas Highland in Namibia at 1800\,m above sea level. The array went in operation in 2004. It is composed of four 12\,m dish diameter telescopes and, since 2012, a central telescope named CT5 with a diameter of 28\,m. 
When operated in the initial configuration of four telescopes, the system has  an energy threshold of $\sim$100\,GeV with an angular resolution better than 0.1$^{\circ}$ and an energy resolution below 15\%. The data analysis and the quality selection are  described in \cite{Aharonian06}. These performances will be boosted by a major upgrade of the cameras finished in early 2017 \cite{Giavitto15}. The performances of CT5, currently operated in mono mode, are explained in detail in \cite{HESSmono}. 

MAGIC \cite{magicperf_1:2015}, at the Roque de los Muchachos Observatory in the Canary island of La Palma, Spain, is a system of two   telescopes operating since 2009 at 2200\,m above sea level. The single telescope MAGIC-I operated in mono mode from 2004 to 2009 and was renewed in 2011/2012 to match the performances of the newer MAGIC-II.
With their 17\,m diameter dishes, the MAGIC telescopes reach an energy threshold as low as 50\,GeV. The angular resolution, above 200\,GeV  is $<$ 0.07$^{\circ}$, while the energy resolution is 16\% \cite{magicperf_2:2015}. 
The duty cycle of the system has recently been increased thanks to the possibility of taking data under  partial moonlight \cite{magicperf_moon:2017}.

VERITAS \cite{Holder06} is a ground-based gamma-ray instrument operating at the Fred Lawrence Whipple Observatory (FLWO) in southern Arizona, USA. VERITAS is designed to measure gamma rays with energies from $\sim$\,85 GeV up to $>$ 30 TeV. Since its first light observation in 2007, VERITAS underwent two major upgrades: the relocation of the first telescope in 2009 and the upgrade of the cameras in 2012. VERITAS, together with MAGIC,  observes sources located in the Northern emisphere, while H.E.S.S. observes the Southern sky.

\begin{figure}
\centerline{\includegraphics[width=0.95\linewidth]{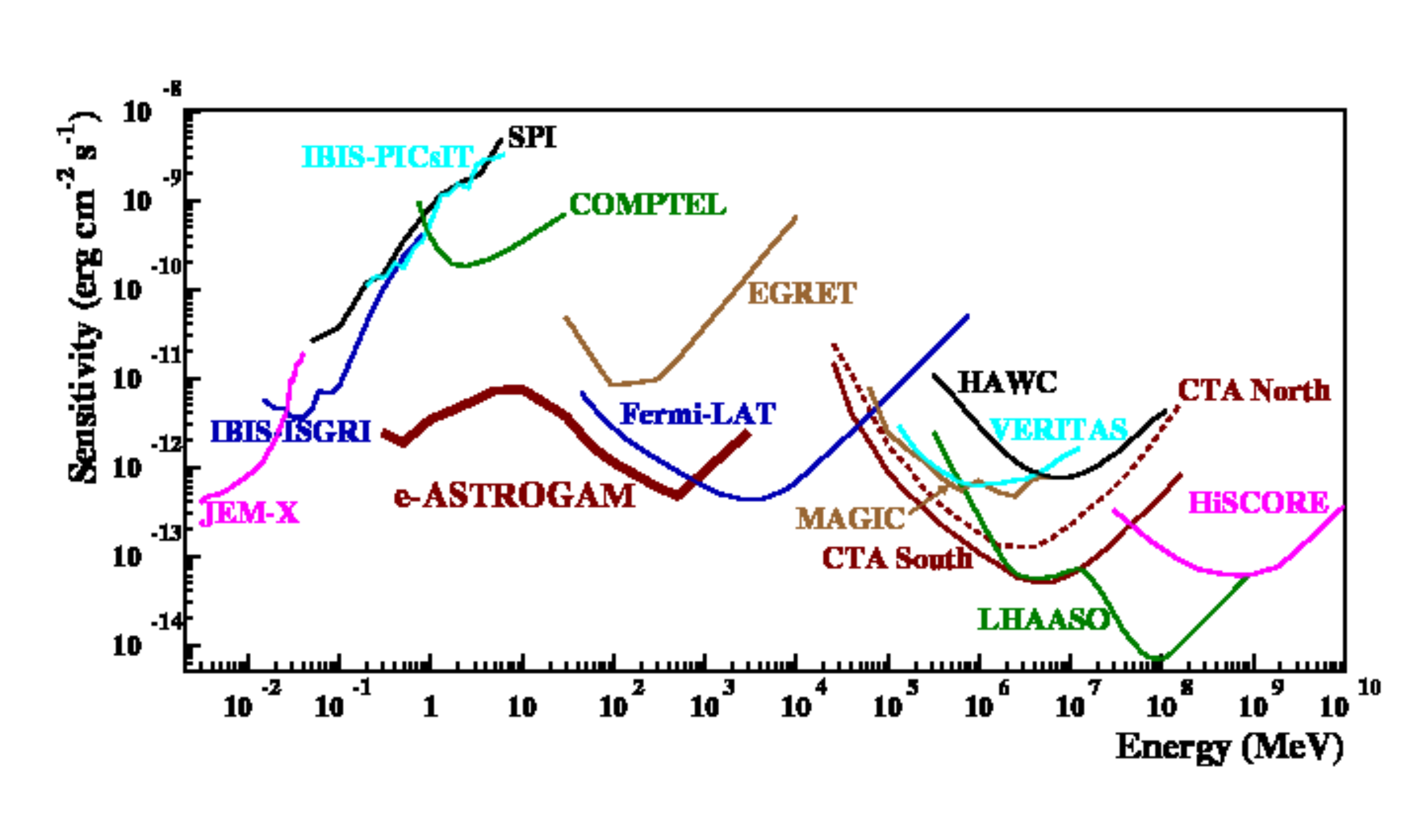}}
\caption[Differential sensitivity of MAGIC and VERITAS telescope systems.]{Differential sensitivity of MAGIC and VERITAS telescope systems in comparison with other telescopes operating or planning to operate in neighboring energy ranges. From \cite{deangelis16}.}
\label{fig:sensitivity}
\end{figure}

Figure~\ref{fig:sensitivity}, from \cite{deangelis16}, shows the differential sensitivity of the VERITAS and MAGIC telescopes reached with 50 hours of obervations in comparison with the sensitivity of other telescopes operating or planning to operate (e.g. LHAASO and e-ASTROGAM) in neighboring energy ranges. 
The figure reports also the expected sensitivity of the future Cherenkov Telescope Array (CTA) \cite{Actis11}, designed to bring the VHE gamma ray astrophysics in the regime of large numbers ($>$ 1000 sources). CTA will consist of two sites, one in the Northern and one in the Southern emisphere, and tens of telescopes of three different sizes. Currently several prototypes are under construction and test. CTA is expected to start its early science in few years from now. See \cite{2017arXiv170507805M} for a review.

The duty cycle of current IACTs is mainly limited by sky brigthness and weather conditions and is below 10\%. 
For this reason, and for the fact that the field of view is small (3.5 - 5$^{\circ}$) forcing IACTs to operate mostly in pointing mode, the H.E.S.S., MAGIC, and VERITAS collaborations have developed a dense Target of Opportunity (ToO) program. 
ToO observations are usually triggered by a high state of the source detected in optical and/or X-ray and/or gamma rays. Additionally, triggers by neutrino and gravitational waves are also considered. 
This strategy allowed the discovery of several sources during flaring states and is particularly relevant for distant emitters and for sources with a faint emission at TeV energies during the quiescent state. 

In this report I will first  make an overview of the TeV extragalactic sky, showing that the quasi totality of the VHE gamma-ray emitters are active galactic nuclei (AGN).
I will then introduce the current paradigm explaining the different flavours of jetted AGNs with a particular emphasis on the blazar sequence, recently revised. After this introduction I will start the detailed report on recent highlight results from the current IACTs, with three sections dedicated to different TeV emitters, and a last section focused on cosmology and propagation studies with VHE gamma rays.
I will conclude with a short summary and a list of future perspectives of this exciting and very fecund research field.

\begin{table}[bht]
\tiny
\caption[]{Updated list of TeV extragalactic sources (Adapted from TeVCat\footnote{Website: http://tevcat.uchicago.edu}).}
\label{tab:src}
\vspace{0.4cm}
\begin{center}
\begin{tabular}{|l|c|c|c|c|}
\hline
{\bf Source Name} & {\bf R.A.} & {\bf Dec} & {\bf Class} & {\bf Distance $z$}\\ 
\hline
S3 0218+35	&	02 21 05.5	&	+35 56 14	&	FSRQ	&	0.95	\\
PKS 1441+25	&	14 43 56.9	&	+25 01 44	&	FSRQ	&	0.94	\\
3C 279    	&	12 56 11.1	&	-05 47 22	&	FSRQ	&	0.54	\\
PG 1553+113	&	15 55 44.7	&	+11 11 41	&	HBL	&	$>$ 0.4	\\
1ES 0033+595	&	00 35 16.8	&	+59 47 24.0	&	HBL	&	0.47	\\
1ES 0647+250	&	06 50 46.5	&	+25 03 00	&	HBL	&	0.45	\\
PKS~1222+216	&	12 24 54.4	&	+21 22 46	&	FSRQ	&	0.43	\\
PKS 1510-089	&	15 12 52.2	&	-09 06 21.6	&	FSRQ	&	0.36	\\
PKS 0447-439	&	04 49 28.2	&	-43 50 12	&	HBL	&	0.34	\\
1ES 0502+675	&	05 07 56.2	&	+67 37 24	&	HBL	&	0.34	\\
OT 081	        &	17 51 32.82	&	+09 39 00.73	&	LBL	&	0.32	\\
S5 0716+714	&	07 21 53.4	&	+71 20 36	&	IBL	&	0.31	\\
OJ 287	        &	08 54 48.9	&	+20 06 31	&	BL Lac?	&	0.3	\\
1ES 0414+009	&	04 16 52.96	&	+01 05 20.4	&	HBL	&	0.29	\\
PKS 0301-243	&	03 03 23.49	&	-24 07 35.86	&	HBL	&	0.26	\\
MS 1221.8+2452	&	12 24 24.2	&	+24 36 24	&	HBL	&	0.22	\\
1ES 1011+496	&	10 15 04.1	&	+49 26 01	&	HBL	&	0.21	\\
RBS 0723	&	08 47 12.9	&	+11 33 50	&	HBL	&	0.2	\\
RBS 0413	&	03 19 47	&	+18 45 42	&	HBL	&	0.19	\\
PKS 0736+017	&	07 39 18.0	&	+01 37 05	&	FSRQ	&	0.19	\\
1ES 0347-121	&	03 49 23.0	&	-11 58 38	&	HBL	&	0.19	\\
1ES 1101-232	&	11 03 36.5	&	-23 29 45	&	HBL	&	0.19	\\
1ES 1218+304	&	12 21 26.3	&	+30 11 29	&	HBL	&	0.18	\\
RX J0648.7+1516	&	06 48 45.6	&	+15 16 12	&	HBL	&	0.18	\\
H 2356-309	&	23 59 09.42	&	-30 37 22.7	&	HBL	&	0.16	\\
1RXS J101015.9-311909	&	10 10 15.03	&	-31 18 18.4	&	HBL	&	0.14	\\
1ES 0229+200	&	02 32 53.2	&	+20 16 21	&	HBL	&	0.14	\\
1ES 0806+524	&	08 09 59	&	+52 19 00	&	HBL	&	0.14	\\
S3 1227+25	&	12 30 14.1	&	+25 18 07	&	IBL	&	0.13	\\
RX J1136.5+6737	&	11 36 30.1	&	+67 37 04	&	HBL	&	0.13	\\
1ES 1215+303	&	12 17 48.5	&	+30 06 06	&	HBL	&	0.13	\\
H 1426+428	&	14 28 32.6	&	+42 40 21	&	HBL	&	0.13	\\
RGB J0710+591	&	07 10 26.4	&	59 09 00	&	HBL	&	0.12	\\
B3 2247+381	&	22 50 06.6	&	+38 25 58	&	HBL	&	0.12	\\
PKS 2155-304	&	21 58 52.7	&	-30 13 18	&	HBL	&	0.12	\\
VER J0521+211	&	05 21 45	&	+21 12 51.4	&	IBL	&	0.11	\\
1ES 1312-423	&	13 14 58.5	&	-42 35 49	&	HBL	&	0.1	\\
W Comae	        &	12 21 31.7	&	+28 13 59	&	IBL	&	0.1	\\
SHBL J001355.9-185406	&	00 13 52.0	&	-18 53 29	&	HBL	&	0.09	\\
1ES 1741+196	&	17 44 01.2	&	+19 32 47	&	HBL	&	0.08	\\
RGB J0152+017	&	01 52 33.5	&	+01 46 40.3	&	HBL	&	0.08	\\
PKS 2005-489	&	20 09 27.0	&	-48 49 52	&	HBL	&	0.07	\\
BL Lacertae	&	22 02 43.3	&	+42 16 40	&	IBL	&	0.07	\\
PKS 0548-322	&	05 50 38.4	&	-32 16 12.9	&	HBL	&	0.07	\\
PKS 1440-389	&	14 43 57.2	&	-39 08 40	&	HBL	&	0.06	\\
1ES 1727+502	&	17 28 18.6	&	+50 13 10	&	HBL	&	0.05	\\
PKS 0625-35	&	06 27 06.7	&	-35 29 15	&	FRI	&	0.05	\\
1ES 2037+521	&	20 39 23.5	&	52 19 50	&	HBL	&	0.05	\\
AP Librae	&	15 17 41.8	&	-24 22 19	&	LBL	&	0.05	\\
1ES 1959+650	&	19 59 59.8	&	+65 08 55	&	HBL	&	0.05	\\
Markarian 180	&	11 36 26.4	&	+70 09 27	&	HBL	&	0.04	\\
1ES 2344+514	&	23 47 04.9	&	+51 42 17	&	HBL	&	0.04	\\
Markarian 501	&	16 53 52.2	&	+39 45 37	&	HBL	&	0.03	\\
Markarian 421	&	11 04 19	&	+38 11 41	&	HBL	&	0.03	\\
IC 310	        &	03 16 43.0	&	+41 19 29	&	HBL	&	0.02	\\
NGC 1275	&	03 19 48.1	&	+41 30 42	&	FRI	&	0.02	\\
M 87	        &	12 30 47.2	&	+12 23 51	&	FRI	&	0.004	\\
Centaurus A	&	13 25 26.4	&	-43 00 42	&	FRI	&	0.002	\\
M 82	        &	09 55 52.7	&	+69 40 46	&	Starburst	&	3900 kpc	\\
NGC 253	        &	00 47 06	&	-25 18 35	&	Starburst	&	2500 kpc	\\
1ES 1440+122	&	14 43 15	&	+12 00 11	&	HBL	&	-	\\
RGB J2056+496	&	20 56 42.7	&	+49 40 07	&	Blazar	&	-	\\
S2 0109+22	&	01 12 05.8	&	+22 44 39	&	IBL	&	-	\\
S4 0954+65	&	09 58 47.00	&	65 33 55.00	&	Blazar	&	-	\\
RGB J2243+203	&	22 43 54.7	&	+20 21 04	&	HBL	&	-	\\
H 1722+119	&	17 25 04.3	&	11 52 15	&	HBL	&	-	\\
RGB J0136+391	&	01 36 32.5	&	+39 06 00	&	HBL	&	-	\\
KUV 00311-1938	&	00 33 34.2	&	-19 21 33	&	HBL	&	-	\\
HESS J1943+213	&	19 43 55	&	+21 18 08	&	HBL	&	-	\\
PKS 1424+240	&	14 27 00	&	+23 47 40	&	HBL	&	-	\\
\hline
\end{tabular}
\end{center}
\end{table}

\section{An overview of the TeV extragalactic sky}\label{sec:TeVsky}

\begin{figure}
\centerline{\includegraphics[width=0.6\linewidth]{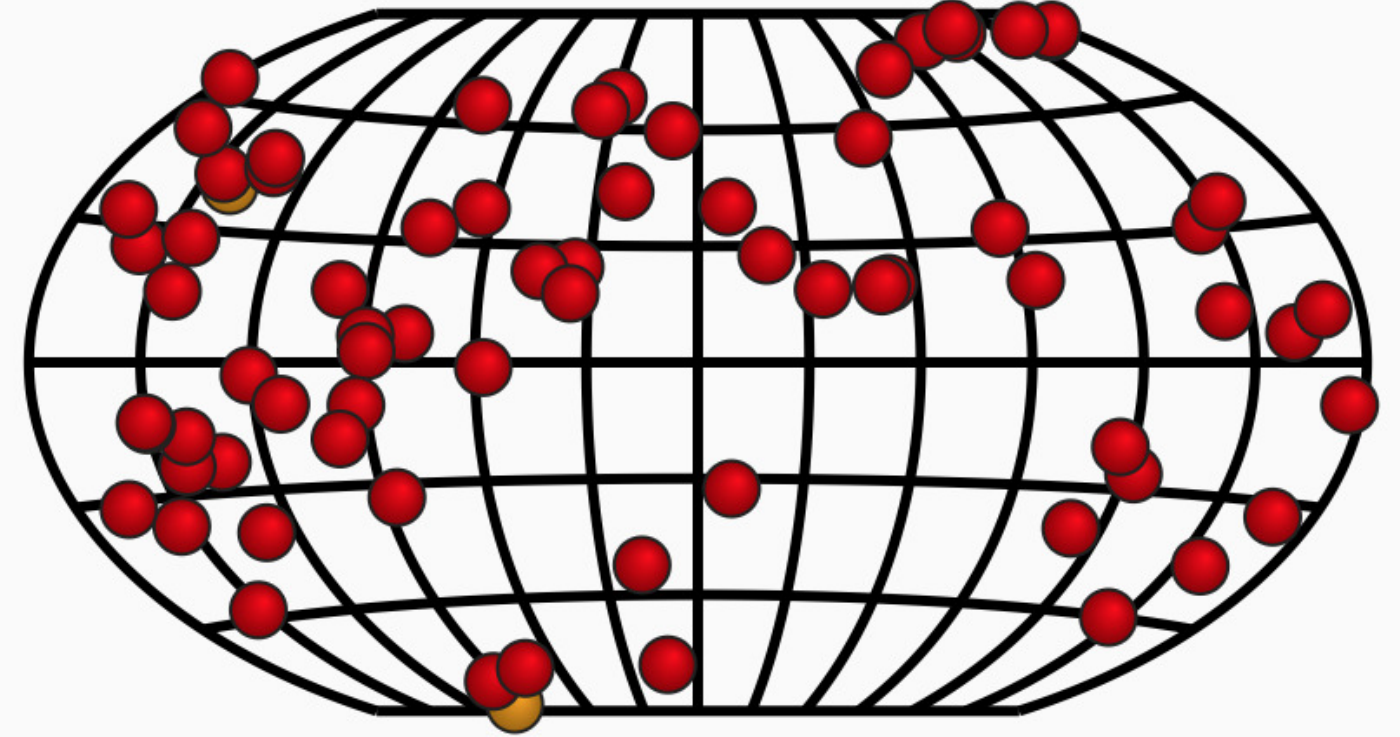}}
\caption{Map of the TeV-detected extragalactic sources. Red filled dots are AGNs, while the two orange filled dots represent starburst galaxies.}
\label{fig:TeVsky}
\end{figure}

A list of TeV-emitting extragalactic sources detected by current generation of IACTs is reported in Table~\ref{tab:src} and plotted in Figure~\ref{fig:TeVsky}. The main features characterizing the detected sources are:
\begin{itemize}
\item The large majority of the TeV extragalactic emitters are classified as high synchrotron peaked BL Lac objects (see next section for details); 
\item The only non-AGN sources are two nearby starburst galaxies: M~82 and NGC~253 (orange filled dots in the Figure);
\item The redshift is limited below 1, with the large majority of the sources featuring a redshift below 0.5;
\item A number of detected blazars have no redshift measurement. This is a quite typical feature of BL Lac objects, that may have no emission/absorption lines and whose optical non-thermal continuum is so strong to prevent the detection of the host galaxy.
\end{itemize}

\section{Jetted AGNs and the Blazar Sequence}\label{sec:AGNmodel}
An AGN is defined as a compact region at the center of a galaxy showing a luminosity much higher than normal, mostly of non-stellar origin.
A plethora of different AGN categories have been proposed in literature starting from the early seventies, according to some spectral features such as the characteristics of emission and absorption lines in the spectrum or the radio luminosity. The latter criterion separates the sample into the two main classes  of radio-loud and radio-quiet AGNs.

In 1995, after several years of astronomical discoveries, Urry and Padovani proposed the innovative idea that different features seen in the spectra of radio-loud AGNs could be related to the different orientation of the same, complex,  object lying in the nucleus of the active galaxies \cite{Urry95}. Since then, radio loud AGNs have been described as supermassive black holes accreting material and ejecting part of it through two collimated jets of particles accelerated to ultra-relativistic energies (in radio quiet objects the structure is similar but there are no jets). 
Very recently, in order to include a sample of radio-faint objects showing many characteristics in common with radio-loud AGNs \cite{2011RAA....11.1266F}, the new definition of {\it jetted AGNs} instead of radio-loud AGNs has been proposed \cite{Padovani16}, and will be used in what follows.

Jetted AGNs usually show a double-peaked SED covering almost all the electromagnetic spectrum, from radio to gamma-ray frequencies. The first peak of the SED is due to synchrotron radiation by accelerated electrons in the jet, and the second is most likely related to inverse Compton emission. Additional components in the SED related to the light emitted by the accretion disc or a hot corona may be present. For a recent review see \cite{galaxies4040037}.

The two main classes of jetted AGNs are:
\begin{description}
\item[Radiogalaxies]: when the viewing angle connecting the observer with the jet direction is relatively large ($>$20$^{\circ}$). In radio, these objecs usually show two thin radio structures extending up to several kpc and identified as the emission from the two jets. 
\item[Blazars]:  when the viewing angle connecting the observer with the jet direction is relatively small ($<$ 20$^{\circ}$). Due to ultra-relativistic velocities of the particles in the jet, the jet emission is stronlgy boosted and dominates the overall luminosity. Depending on the presence of  emission lines in the spectrum, blazars are further divided into Flat Spectrum Radio Quasars (FSRQs) and BL Lac objects, with the former characterized by strong broad emission lines in the optical spectrum. 
\end{description}

A very interesting feature of blazars SED is the anticorrelation between their bolometric luminosity and the position of the synchrotron peak, the so-called blazar sequence \cite{Fossati98}. A recent study performed by Ghisellini \cite{Ghisellini17} explores the same sequence in a subsample of $\sim$750 blazars detected by {\it Fermi}-LAT, and confirms that also the second peak of the SED follows the same trend (even if less pronounced). FSRQs show a very bright spectrum with the SED peaks located mainly towards low energies while BL Lacs feature a lower luminosity and the peaks shifted at higher energies. The subclasses of low/intermediate/and high synchrotron peaked BL Lac objects (LBL, IBL, and HBL respectively) seem to reflect this behaviour. 
Looking at this sequence, it is not surprising that the large majority of blazars detected by IACTs are HBLs: blazars with a synchrotron and a high energy peak shifted towards the highest energies, while FSRQs are usually detected at VHE only during flaring episodes.

After almost fifty years of jetted-AGN observations, we have now many experimental evidences supporting a general picture, but also many unknowns. These are: the size and location of the emitting region, the role of hadrons in the jet, and the acceleration mechanism at work (shocks or other processes). 

Current IACTs are optimizing the observation time in order to  discover new sources of VHE radiation and study in detail the emission from known sources. In both cases, the collaboration with telescopes observing in other bands is crucial. New discoveries are in fact very often triggered by an optical or gamma-ray high-state of a good candidate, while detailed studies are carried out in a multi-wavelength approach. 




\subsection{The Gamma-ray horizon}
The limited redshift distribution that characterizes the TeV-detected blazars, if compared for example to the blazars detected at lower frequencies by {\it Fermi}-LAT satellite \cite{3LAC}, can only in part be attributed to the extreme energetics needed to produce TeV gamma rays. The other reason is  related to the interaction of VHE photons with the optical and infrared photons filling the Universe, the so-called extragalactic background light (EBL) \cite{Stecker92}.
EBL is the sum of the light emitted by the stars and reprocessed by the dust since the birth of the first stars. When a TeV gamma ray is produced, it may interact with EBL photons and create an electron/positron pair. This process depends on the energy of the photon and on the distance of the emitter: in closeby sources only the most energetic photons are affected (above few TeV), while at larger redshift ($>$0.3) this process is already effective at a few hundreds GeV. For sources located at redshift 1, 100\,GeV photons are already significantly absorbed.

Different EBL models have been developed, which try to describe the EBL energy density and its evolution with cosmic time. Last generation of models, e.g. \cite{Franceschini08,Dominguez11}, are pretty in agreement and seem to suggest that we are close to resolve the quasi totality of the sources emitting optical radiation. 

\section{Gamma rays from misaligned blazars}\label{sec:rgal}

In view of the unified model presented in previous section, radiogalaxies can be considered as {\it misaligned blazars}. Thanks to the large viewing angle, they offer the unique opportunity to localize the emitting region of blazars. 
Four of these objects have been up to now detected up to the extreme energies, namely PKS~0625--35, M~87,  NGC~1275, and Cen~A. Additionally, IC~310 can be included in this family due to its uncertain nature. 

Interesting results from Cen~A and IC~310 were recently published and I will shortly describe them in the following subsections. The interested reader is addressed to \cite{2016arXiv161102986R} and references therein for an exhaustive description. 

\subsection{Cen~A}
Centaurus~A is the closest and best studied radio galaxy. Its morphology was observed in detail and it is quite complex, showing jets at different scales (sub-pc to kpc), radio lobes, and an extended emission in many bands, including gamma rays.

Usually the broadband SED  from the core of radio galaxies is well described by a synchrotron self-Compton (SSC) model, with the low energy emission due to synchrotron radiation emitted by relativistic electrons, and the high energy component due to the very same electrons upscattering the synchrotron photons via inverse Compton mechanism. 

\begin{figure}
\centering
\includegraphics[width=0.48\linewidth]{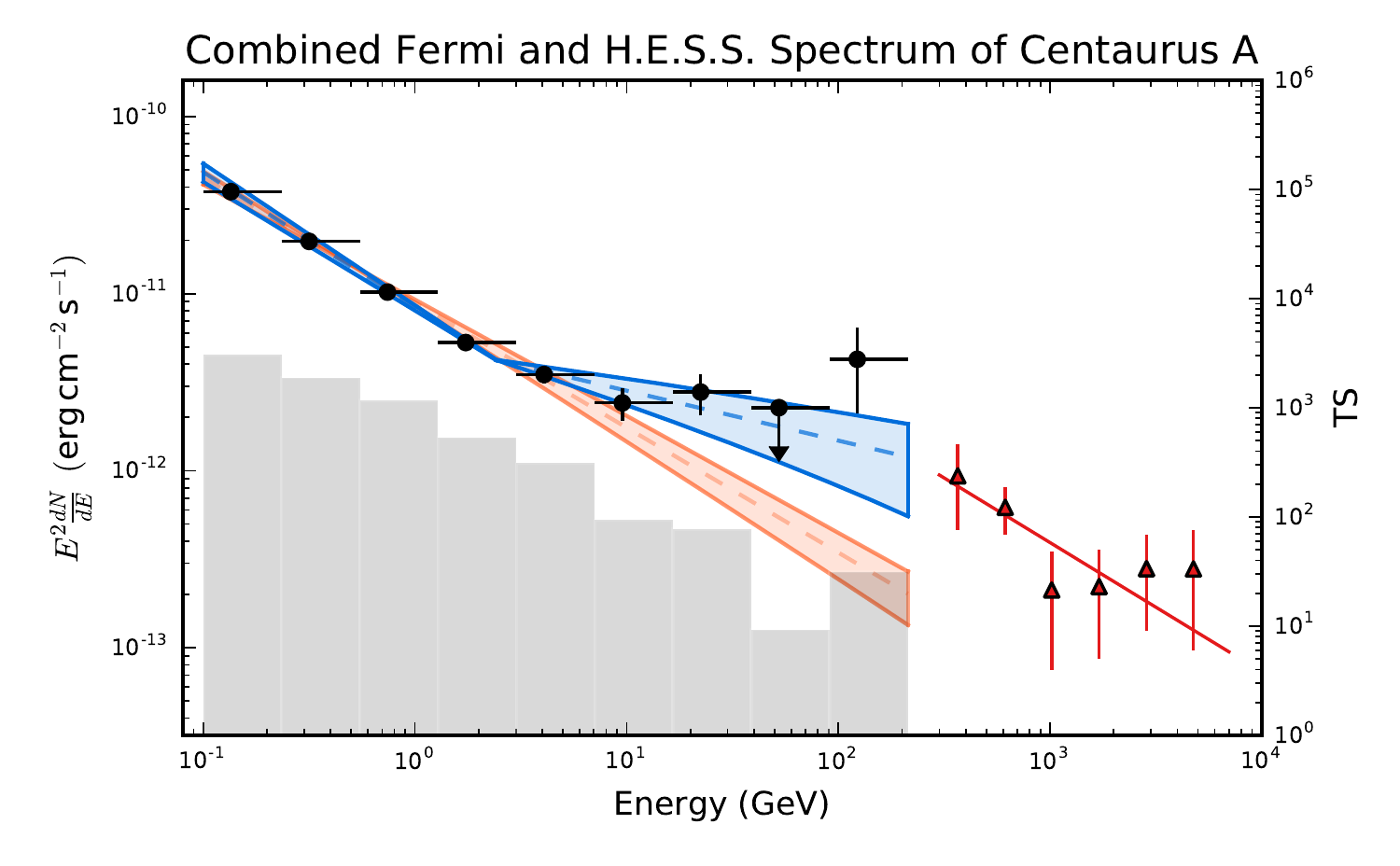}
\includegraphics[width=0.48\linewidth]{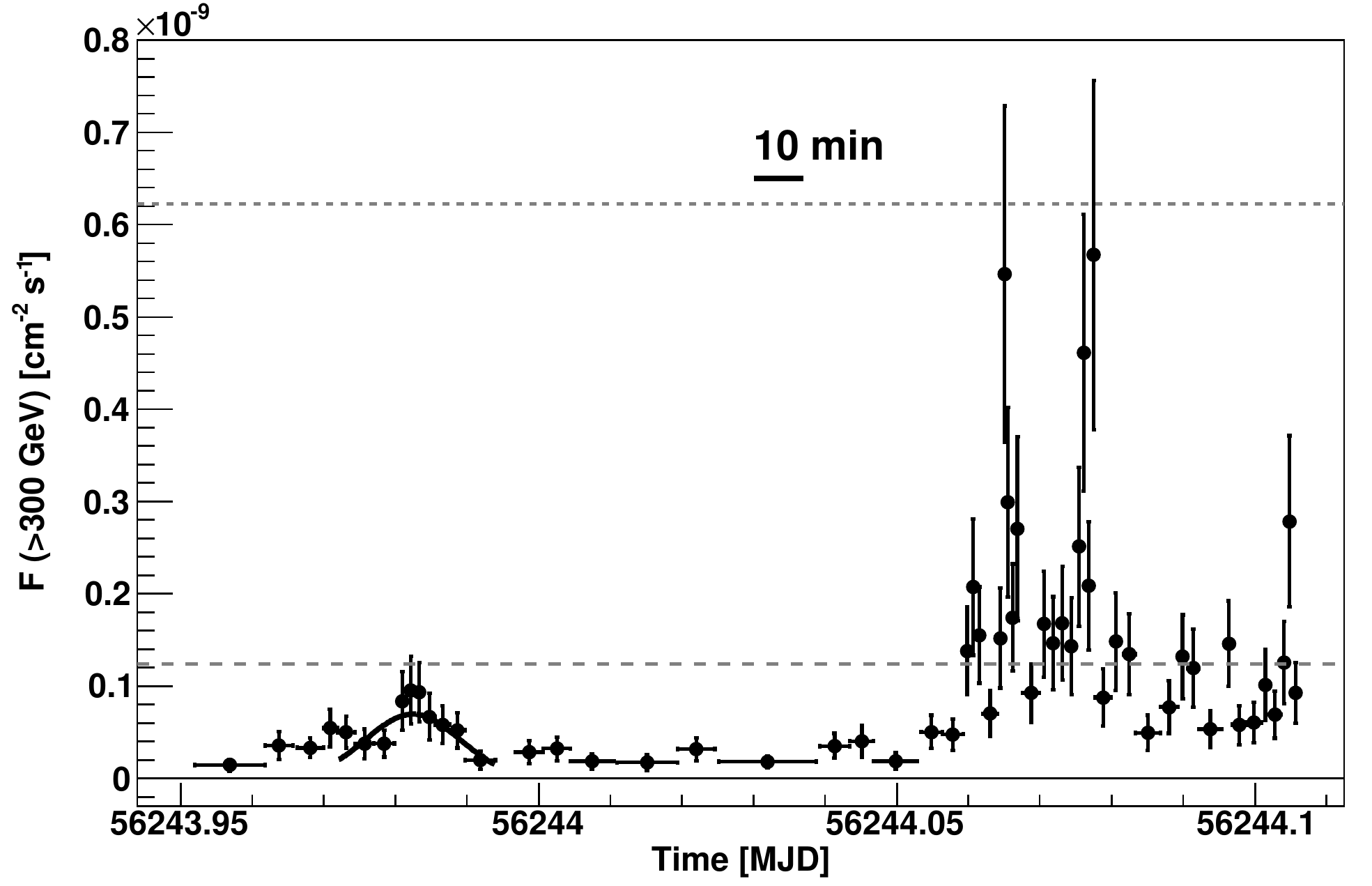}
\caption{Recent findings from misaligned blazars. Left plot: Combined high energy spectrum of Cen~A, from \cite{Brown17}, showing an unexpected hardening above few GeV. Black filled points are {\it Fermi}-LAT data, while red triangles represent the H.E.S.S. spectrum measured at the VHE. Right plot: VHE gamma rays lightcurve of IC~310, exhibiting an extremely fast variability down to 4.8 minute timescale. From \cite{Aleksic14}.}
\label{fig:MisalignedBzb}
\end{figure}

The  detection of an unexpected hardening in the high energy spectrum from 7.5 years of {\it Fermi}-LAT data in the core of Cen~A was recently published in \cite{Brown17}.
Left panel of Figure~\ref{fig:MisalignedBzb}, from \cite{Brown17}, shows the spectrum measured with {\it Fermi}-LAT (black dots) together with the long-term VHE spectrum measured with H.E.S.S. (red triangles).
Interestinlgy, below few GeV the spectral slope is compatible with the slope found at the highest energies, which are however shifted toward a higher flux density. The spectral hardening detected in {\it Fermi} data well connects these two regimes and strongly supports the hypothesis of an {\it additional emission component in the spectrum of Cen~A} extending at VHE. According to the authors, this component could be due to a spike in the dark matter profile, to a population of millisecond pulsars, or to an additional hadronic component. 

\subsection{IC 310}

Another uncommon feature that has been recently detected in a misaligned blazar is the extremely short timescale variability seen in IC~310, a known TeV emitter located in the Perseus cluster of galaxies.

The MAGIC telescopes observed  IC~310 in November 12-13, 2012 \cite{Aleksic14}. The light curve measured above 300\,GeV, reported in right panel of Figure~\ref{fig:MisalignedBzb}, revealed an incredibly fast variability with a flux doubling time shorter than 5 minutes. 
This measure allowed a very constraining measurement of the size of the emitting region. Considering that causality forces this size {\it s} to be smaller than $c\,\delta t$, $s$ results smaller than 20\% of the gravitational radius of the central black hole.

Also in this case, this experimental evidence marks a break of the standard paradigm of emission from the nucleus of a jetted active galaxy. According to the standard SSC model, in fact, the size of the emitting region cannot be smaller than the jet width, as in IC~310. 
The most natural explanation of this finding is the  occurrence of vacuum gap-type particle acceleration on sub-horizon scales, a mechanism often invoked to explain pulsar emission.

\section{Gamma rays from FSRQs}\label{sec:fsrq}
FSRQs are the majority of extragalactic sources detected in the high energy range by {\it Fermi}-LAT, while they constitute just a small fraction of the TeV-detected sources. As mentioned above, this evidence can be fully explained with the blazar sequence: FSRQs are very bright objects whose SED high-energy peak is located in the sub-GeV range. Moreover, the typical redshift of FSRQs is well above 1, making them difficult targets for IACTs due to the interaction with EBL.

In the last few years, VHE observations have challenged the standard emission model proposed for these objects, which explained the observed non-thermal continuum as syncrotron radiation (at low energies) and inverse Compton emission on external photon fields (also called {\it external Compton}, at higher energies). In FSRQs, the presence of an internal region close to the central object characterized by a dense photon field (the {\it broad line region, BLR}), implies that if VHE radiation is emitted within this region, it must be self-absorbed. 
The issue arised when an unexpected short variability timescale was observed at VHE from the FSRQs PKS~1222+216 \cite{Aleksic11}. Causality forces the emitting region to be very compact, meaning close to the black hole if we assume the canonical shock-in-the jet model. But no VHE emission should escape this region due to self absorption, as mentioned above.

Seven FSRQs have been up to now detected at the highest energies. Two of them are located at surprisingly high redshift: PKS 1441+25 at $z = 0.939$ and B0218+357 at $z = 0.944$, and will be briefly outlined.

\subsection{PKS~1441+25}

\begin{figure}[b]
 \begin{minipage}[c]{0.7\textwidth}
   \includegraphics[width=0.9\textwidth]{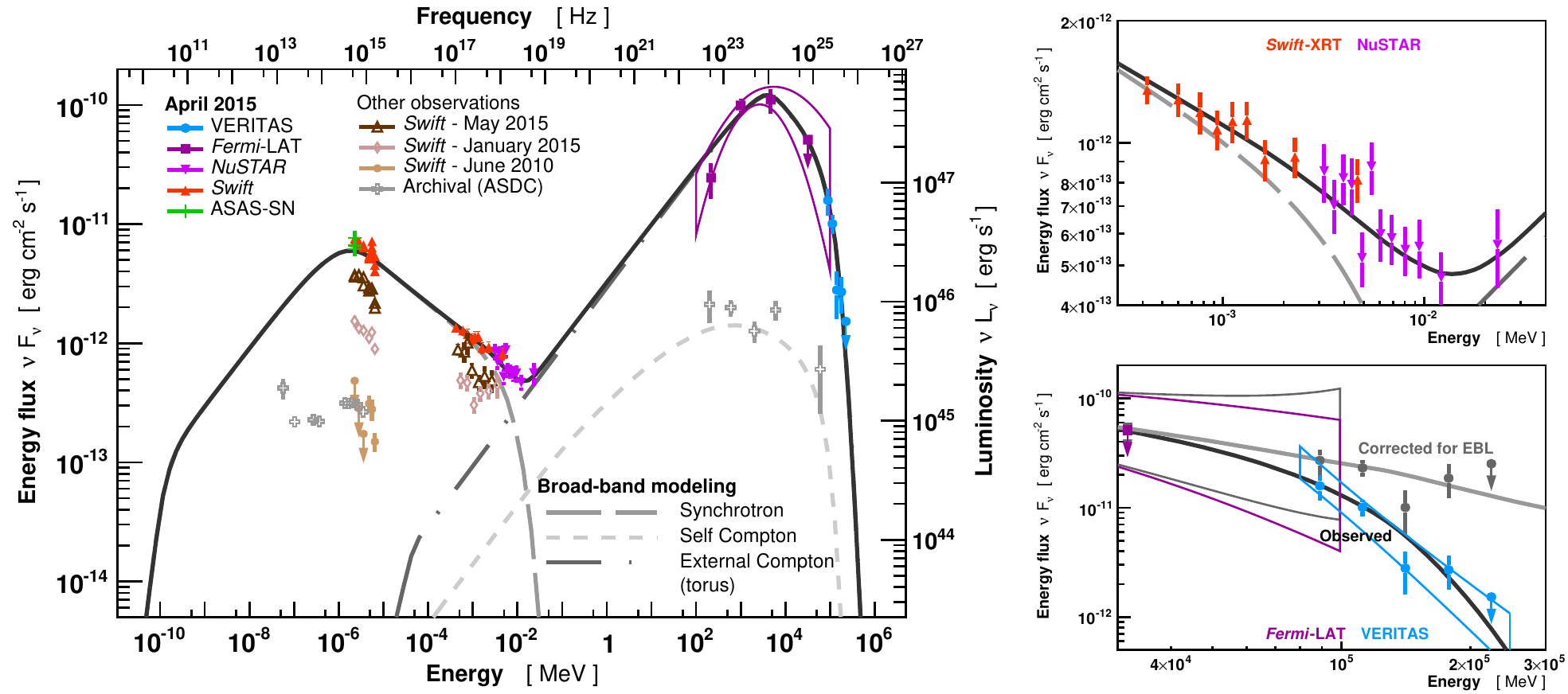}
 \end{minipage}\hfill
 \begin{minipage}[c]{0.3\textwidth}
   \caption{Recent findings from VHE gamma-ray observations of FSRQs. Left plot: multi-frequency SED of PKS~1441+25 from quasi-simultaneous data in April 2015 \cite{2015ApJ...815L..22A}. The SED is well fitted with the SSC plus external Compton model.}\label{fig:FSRQs}
 \end{minipage}
\end{figure}

A significant VHE signal from PKS~1441+25 was detected in the dataset collected with the MAGIC telescopes the second half of April 2015 \cite{2015ApJ...815L..23A}. The observations were triggered by a high emission state seen with {\it Fermi}-LAT. VERITAS also observed the source and detected a significant signal \cite{2015ApJ...815L..22A}.

A dense multi-wavelength campaign on the source was carried on during this episode. The large distance of PKS~1441+25 implies a very strong absorption of the VHE spectrum already above 100\,GeV. Very remarkably, once corrected for the EBL aborption the spectrum connects smoothly with the simultaneous spectrum measured by {\it Fermi}-LAT and the overall SED is nicely fitted with a SSC model with an additional component due to external Compton of infra-red photons coming from a dusty torus sourrunding the accretion disk, see left plot in Figure~\ref{fig:FSRQs}. The variability timescale is constrained to few days, and is in agreement with the shock-in-the-jet model once assumed that the emitting region lies outside the BLR.

\subsection{B0218+357}

\begin{figure}
 \begin{minipage}[c]{0.7\textwidth}
   \centering
   \includegraphics[width=0.9\textwidth]{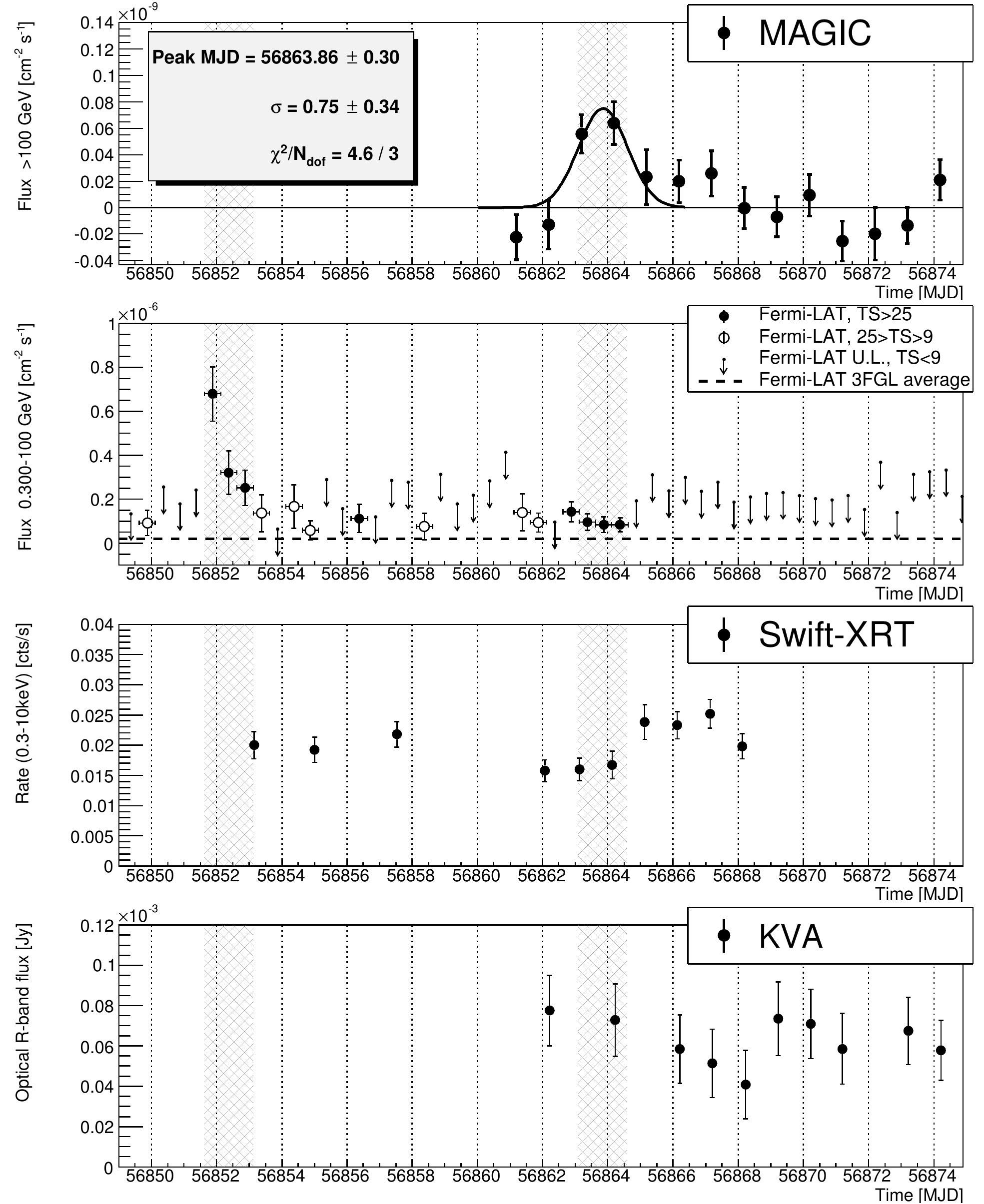}
 \end{minipage}\hfill
 \begin{minipage}[c]{0.3\textwidth}
   \caption{Multi-band light curve of the gravitationally lensed FSRQ B0218+357. A clear TeV emission from the delayed component was detected by the MAGIC telescopes. From  \cite{2016AA...595A..98A}.}\label{fig:FSRQs_0218}
 \end{minipage}
\end{figure}

Two main features characterize the FSRQ B0218+357: its large redshift, making this source the most distant TeV emitter known to date, and the fact that this source is a gravitationally lensed blazar. 
The lens is the spiral galaxy B0218+357G located at a redshift of $z~\sim 0.68$. The radio image clearly shows two components, with a delay between the two images measured during flaring episodes of 10-12 days. This delay was also seen in 2012 at high energies by {\it Fermi}-LAT during a series of outbursts \cite{2014ApJ...782L..14C}. 

The successive strong flare from the source was registered by {\it Fermi} in 2014, and triggered observations with the MAGIC telescopes. However MAGIC could point the source only during the expected arrival time of the delayed component of the emission. A significant signal was detected by MAGIC below 200~GeV \cite{2016AA...595A..98A}. Figure~\ref{fig:FSRQs_0218} shows the VHE and high-energy gamma rays, X-ray, and optical light curve of this event. 
Interestingly, two distinct emitting region in the jet are necessary to explain the observed SED, which is well reproduced in the framework of a two zones external Compton scenario, as detailed in  \cite{2016AA...595A..98A}.

\section{Gamma rays from BL Lac objects}\label{sec:blac}
BL Lacs form the most numerous class of extragalactic objects seen at VHE gamma rays. In the last five years, IACTs discovered more than 10 new sources of this class, mainly thanks to high-energy gamma-, X-ray and optical alerts (e.g. the very recent detection of the blazar OJ~287 reported by the VERITAS team in the ATel \#10051).

Moreover, a number of known emitters are monitored regularly with quasi simultanoeus data collected in different bands. These multi-wavelength campaigns are aimed at characterizing the physical properties of the jet. An example of monitoring during a quiescent state of Mkn~421 is presented in  \cite{2016ApJ...819..156B}, where MAGIC and VERITAS data are combined for the first time with NuSTAR  (hard X-ray band) and many other data from radio to gamma rays to build and model a very detailed multi-epoch SED.
Known sources are often use to calibrate instruments, like the case of PG~1553+113.

\subsection{PG~1553+113}
PG~1553+113 is one of the few sources that can be seen by all the three main IACTs currently in operation. It is a BL Lac with unknown redshift, most likely lying between 0.43 and 0.58 \cite{2010ApJ...720..976D}.
Currently PG~1553+113 is the object of a deep, regular multi-wavelength campaign at several bands including VHE gamma rays. This campaign followed the claim of a quasi-periodicity in the gamma ray and optical lightcurves seen by the {\it Fermi}-LAT team \cite{2015ApJ...813L..41A}.

\begin{figure}
\centering
\includegraphics[width=0.7\linewidth]{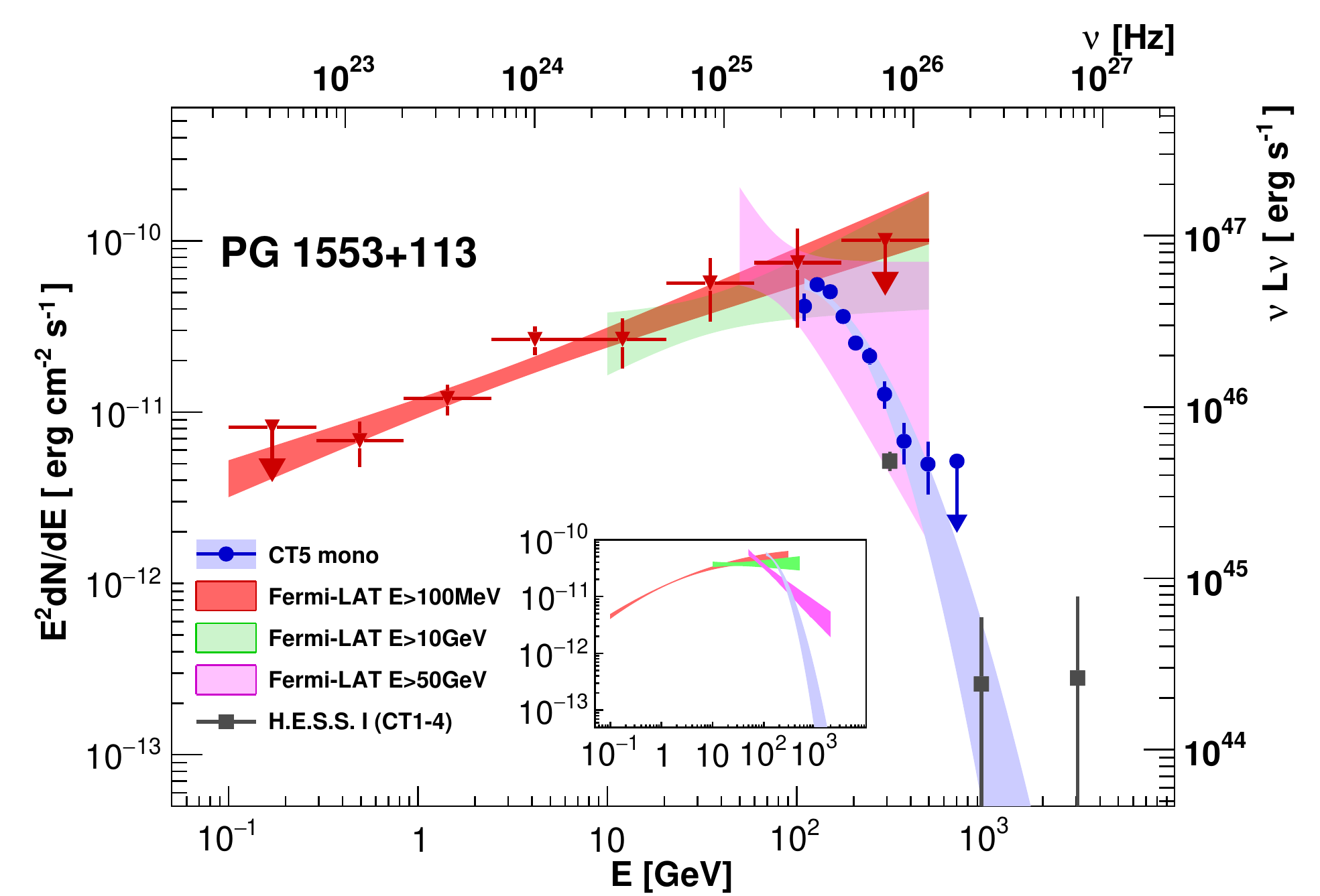}
\caption{SED of PG~1553+113 as measured by CT5 (blue dots) and {\it Fermi}-LAT (red triangles) simultaneously. From \cite{2017AA...600A..89H}.}
\label{fig:bllac}
\end{figure}

Together with PKS~2155-304, PG~1553+113 was used by the H.E.S.S. team to study in detail the performances of the CT5 telescope in mono mode \cite{2017AA...600A..89H}.
Figure~\ref{fig:bllac} shows the gamma-ray SED of the source measured with simultanoeus CT5 and {\it Fermi}-LAT data. The CT5 spectrum is presented with significant data down to 110\,GeV and connects smoothly with lower energy data measured by the LAT instrument. No significant curvature was seen in the spectrum of PG~1553+113. 

\section{Cosmology and Propagation with VHE gamma rays}
Several studies conducted in the last years have demonstrated that VHE gamma rays from cosmological sources are a very useful tool for studying cosmology and photon propagation. In particular VHE gamma rays can be used to test Lorentz invariance violation models, probe the intergalactic magnetic field, test EBL models and the hypothesis of photon oscillation in axion-like particles. I will shortly report on the last two subjects. For a recent review see \cite{2016arXiv161008245D} and references therein.

\subsection{EBL}
The effect on blazars spectra due to the interaction of VHE photons with EBL photons can be used to probe, and in some cases exclude, the EBL models (e.g. \cite{Franceschini08,Dominguez11}). This can be done once an intrinsic shape for the blazar spectra is assumed. First observations of distant blazars, such the FSRQ 3C~279 located at redshift 0.536 and discovered at VHE by MAGIC in 2007 \cite{2008Sci...320.1752M}, provided  constraints on some EBL models in use ten years ago. Once corrected for the absoption in fact, the instrinsic spectrum showed a significant pile up at the highest energies that is very difficult to reconcile with our knowledge of blazar emission. 

\begin{figure}
  \centering
  \includegraphics[width=0.7\linewidth]{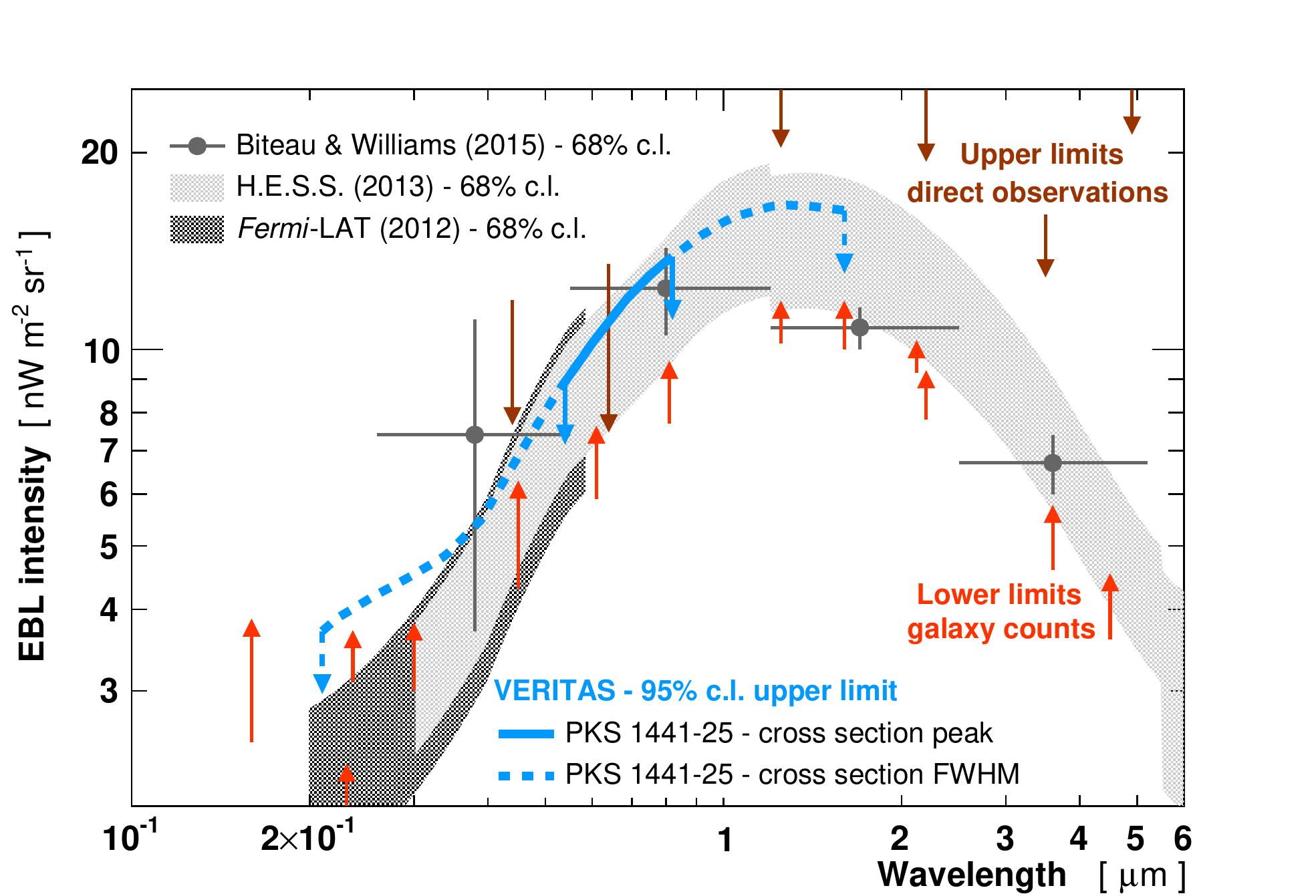}
  \caption{Left: The light blue curve represents the upper limits on the EBL energy density as resulted by VERITAS observations of PKS~1441+25. From \cite{2015ApJ...815L..22A}.}
  \label{fig:cosmo}
\end{figure}

Current models are converging on a similar shape for the EBL. Recent detections of blazars up to redshift $\sim$~1 allowed to probe for the first time new parts of the EBL spectrum (those at the highest energies, in the optical range). Figure~\ref{fig:cosmo} shows the limits provided by VERITAS measurement of the spectrum of PKS~1441+25, already mentioned above \cite{2015ApJ...815L..22A}. The upper limits set are very close to EBL firm lower limits given by direct galaxy counts and are remarkably in agreement with EBL models. 

A multi-source approach have been proposed in several studies \cite{2013AA...550A...4H,2015ApJ...812...60B,2016arXiv161009633M}.
In particular Biteau and Williams used a sample of 106 spectra from 36 different objects and were able to build their own EBL spectrum from mid-IR to far IR. This spectrum is in good agreement with current models and upper/lower limits.

\subsection{Axion-like particle hypothesis}
The long distances travelled by gamma rays into the interstellar space offer the possibility of testing the hypothesis of photons oscillation. A very popular proposal is that VHE gamma rays may oscillate into the so-called axion-like particles (ALP) with a probability related to the external magnetic field, the distance of the emitter, and the $\gamma$-ALP coupling constant (see for example \cite{2009PhRvD..79l3511S}). Hence, this oscillation (including the ALP/$\gamma$-ray reconversion) may happen in the vicinity of the source, in the intergalactic space, or in our galaxy. 

The hypothetical effect of these oscillations is a distorsion of the VHE spectra of an AGN and a partial reduction of the EBL suppression effect, since ALPs do not interact with EBL photons. 
Possible signatures may be wiggles in the VHE spectra of blazars or the observation of appartent high-opacity photons. 

In the last years different studies led to different, in some cases contradictory, results. The analysis of the 106 spectra of blazars by Biteau and Williams resulted into the non detection of any deviation from the expected flux once corrected by EBL effect \cite{2015ApJ...812...60B}. This is in contradiction with the claim done in \cite{2012JCAP...02..033H}, where an evidence for such a deviation is instead found. 

Also in this case, good quality spectra of AGNs, in particular those at  large distances, would be the smoking gun to probe (or discard) this hypothesis.

\section{Summary and Conclusions}
In summary, VHE observations of jetted AGNs provided by IACTs  are important for different aspects. The lightcurves sampled with the good time resolution of  IACTs  are a very powerful tool to constrain the size of the emitting region. Spectral measurements allow to probe the highest energies reached in the jet and test the emission mechanism at work. Moreover, VHE photons, interacting with the EBL, can be used to test EBL models or the hypothesis of oscillation of photons in axion-like particles.

The last few years have seen a number of important results covering many aspects of the extragalactic VHE gamma ray sky. While some results are in pretty good agreement with current models (such as the EBL absoption effect), some others, in particular those related to fast variability, strongly suggest that the standard paradigm of radiogalaxy/blazar emission should be revised. Further measurements of timing and spectral properties of the broadband emission from TeV AGNs are therefore mandatory. This is the main challenge for the IACTs of the next decade, and in particular for the CTA, which is expected to increase the number of known sources by at least one order of magnitude.


\section*{Acknowledgments}
I would like to thank Michele Doro for many fruitful discussions and helpful comments, Giovanni Busetto, Giovanni La Mura, and Mos\'e Mariotti for the careful reading of this manuscript. I am grateful to the TeVCat team for the very useful catalog available to the community. This work is funded by the Padova University and I.N.F.N. Sez. di Padova.


\end{document}